# Modification of intergrain connectivity, upper critical field anisotropy and critical current density in ion irradiated MgB$_2$ films

S. D. Kaushik[1], Ravi Kumar[2], P. K. Mishra[3], J. E. Giencke[4], C. B. Eom[4], and S. Patnaik[1]

[1]*School of Physical Sciences, Jawaharlal Nehru University, New Delhi 110067, India*

[2]*Inter – University Accelerator Center, P. O. Box 10502, New Delhi 110067, India*

[3]*Bhabha Atomic Research Center, Mumbai 400085, India*

[4]*Applied Superconductivity Center and Department of Material Science and Engineering, University of Wisconsin-Madison, WI53705 USA*

## Abstract

We study the effect of 100 MeV Silicon and 200 MeV Gold ion irradiation on the inter and intra grain properties of superconducting thin films of Magnesium Diboride. Substantial decrease in inter-grain connectivity is observed, depending on irradiation dose and type of ions used. We establish that modification of σ band scattering mechanism, and consequently the upper critical field and anisotropy, depends on the size and directional properties of the extrinsic defects. Post heavy ion irradiation, the upper critical field shows enhancement at defect density that is five orders of magnitude less compared to neutron irradiation. The critical current density however is best improved through light ion irradiation.

PACS numbers(s): 74.70.Ad, 74.25.Op, 74.78.Db

# 1. Introduction

The discovery of superconductivity in Magnesium diboride ($T_c \sim 39$ K) has rekindled the hope for possible large-scale applications of superconductors using cryogen-free technology [1-3]. $MgB_2$ is readily available, commercially viable and is devoid of weak link problems across the grain boundaries [2, 3]. Further, the prospect of achieving optimal upper critical fields by tuning the scattering mechanism of this two-band gap superconductor has attracted considerable attention [4-6]. Several studies have indicated that upper critical field ($H_{c2}$) as high as 60 T is possible in this simple intermetallic compound. However no clear reproducible procedure has been suggested to achieve this enhancement. It is well known that irradiation induced point and columnar defects greatly alter the superconducting properties. While the extended defects contribute to enhanced pinning and consequent increase in critical current density ($J_c$), the point defects generally shorten the electron mean free path leading to increase in $H_{c2}$. In the case of $MgB_2$ however, because of the two band gap features, the physics is further compounded depending on the relative effect of disorder vis á vis the two dimensional $\sigma$ and the isotropic $\pi$ bands [4, 7-10]. Further, while there is general agreement that extrinsic defects modify the $\pi$ intraband scattering mechanism, there is no definite implication for the $\sigma$ band properties. This necessitates a careful study to correlate controlled disorder in $MgB_2$ with electromagnetic characterization.

The reported results from irradiation experiments on $MgB_2$ have provided no clear evidence of formation of columnar defects under heavy ion irradiation [11-16]. Shinde et al. have reported only marginal enhancement in $J_c$ at high fields in 200 MeV Ag irradiated films [11]. This was assigned to agglomerated clusters rather than columnar defects. Olsson et al. used 12 GeV $U^{57+}$ and 1.4 GeV $Au^{32+}$ and have seen some improvement in $J_c$, and $H_{c2}$, yet no evidence of anisotropic pinning after irradiation [12]. Following 5.8 GeV lead ion irradiation Chikumoto et al. have reported decrease in magnetization $J_c$ at low fields possibly due to degradation of grain boundary connectivity [13]. Okayasu et al. also have found no enhancement in $J_c$ either at low fields or at high fields post 2.5 MeV electron irradiation [14]. High fluence neutron irradiation results, on the other hand, have shown decrease in $T_c$ and increase in resistivity $\rho(40 K)$, $H_{c2}$, $J_c$ and the irreversibility field $H^*$ [15]. The recent sequential 2 MeV He ion irradiation suggests that damage is essentially limited to the grains themselves without much effect on the connectivity between the grains [16]. The overall picture that emerges is that irradiation with light ions and heavy ions yield different results on the superconducting properties of $MgB_2$.

In this paper we report a comparative study of 200 MeV $Au^{15+}$ heavy ion and 100 MeV $Si^{8+}$ light ion irradiation at different doses on high quality superconducting $MgB_2$ thin films. It is understood that $Si^{8+}$ irradiation will introduce primarily isotropic point defects where as $Au^{15+}$ will give rise to extended disorder or clusters along the direction of irradiation. Our results indicate that the $H_{c2}$-T phase diagram shows distinctly different characteristics for point defects and extended defects. The drastic suppression of $T_c$ and corresponding increase in $H_{c2}$ in heavy ion irradiated samples can only be explained by alteration in $\sigma$ band scattering mechanism along with changes in $\pi$ band scattering. The suppression of intergrain connectivity is found to be fairly independent of defect density but significantly dependent on size of the extrinsic defects. The critical current density and intragrain flux pinning, however, are best improved with light ion irradiation.

## 2. Experiment

High quality $MgB_2$ films used in this study were prepared by depositing Boron on $Al_2O_3$ 0001 substrate by RF magnetron sputtering. The substrate temperature was 500 ºC. The film was then placed in a Ta tube with Mg lumps, which was then placed in a quartz tube. The tube was evacuated to $\sim 10^{-6}$ Torr and backfilled with 10 Torr of Argon. The sample was then annealed at 900 ºC for 5 hours and allowed to cool to room temperature at the rate of 3 degree per minute. The resulting film is ~200nm thick. Details concerning the x-ray and TEM characterization of similar samples are reported elsewhere [17]. Three pieces of size ~ 2 × 5 mm$^2$ and two pieces of ~ 1.1 × 2.5 mm$^2$ were taken out from the single large piece of the film. Leaving out one pristine piece, 4 other pieces were irradiated with a variety of ions, fluence and energy using the 15 UD pelletron accelerator at the Inter-University Accelerator Center, New Delhi. In the text, Sample A, B, C, D, and E stand for unirradiated, $Au^{15+}$ ($1 \times 10^{12}$ ions/cm$^2$) irradiated, $Au^{15+}$ ($5 \times 10^{12}$ ions/cm$^2$) irradiated, $Si^{8+}$ ($5 \times 10^{12}$ ions/cm$^2$) irradiated, and $Si^{8+}$ ($5 \times 10^{13}$ ions/cm$^2$) irradiated respectively. The energy of the silicon and gold ions were set to 100 MeV and 200 MeV respectively. To irradiate the samples the ion beam was focused to a sharp point of size 1mm and the beam was scanned in the area of 10 mm x 10mm using magnetic scanner that ensured uniform irradiation of the whole sample. The beam current was kept low (0.5 pnA) to avoid the heating effects. The irradiation was carried out at room temperature with the beam at an angle ~5$^0$ from c- axis to avoid channeling effects. Stopping power ($S_e$) and range ($R_p$) of the ions for $MgB_2$ calculated by SRIM simulation software [18] are found to be $S_e$ = 18.08 keV/nm and 2.39 keV/nm and $R_p$ = 19.9 μm and 34.3 μm for 200 MeV $Au^{15+}$ and and 100 MeV $Si^{8+}$ ions respectively. It is important to note that at this energy and fluence, the energy deposited per unit volume of the film is approximately $10^{21}$ keV/cm$^3$ and that is comparable to other studies [15,16].

Standard four-probe technique was used for transport measurements. Contacts were made using 44 gauge copper wires with conducting silver paste. The current used in resistivity measurement was fixed at 500μA. The measurements were carried out using a Cryogenic 8T Cryogen-free magnet in conjunction with a variable temperature insert (VTI). Magnetization $J_c$ for the samples was ascertained using a 5T Quantum Design SQUID magnetometer.

## 3. Results and Discussion

Figure 1 shows the zero field resistivity as a function of temperature for the unirradiated and four irradiated MgB$_2$ thin films. The Inset 1 shows the same data up to T = 40 K. The samples were cooled in zero-field down to 30 K and the data were taken in the warming-up cycle. We note that the onset of superconductivity in the pristine sample occurs around 35. 7 K and 50% transition is achieved in the temperature interval of 0.4 K. The suppression in $T_c$ for the pristine sample is primarily due to impurity in Boron target. When an external field is applied, the intersection points between the transition slope line with the normal state resistivity line and the zero resistivity line correspond to $H_{c2}$ and $H^*$ respectively. This is defined schematically in the inset of Figure 2.

As expected, with higher irradiation dose, the onset of superconducting transition shifts to lower temperatures. While for low dose Au$^{15+}$ (sample B) the $T_c$ is decreased to 34.1 K, for the five times higher dose sample C, there is drastic suppression in $T_c$ (32.5 K). Irradiation with 100 MeV Si$^{8+}$ also shows decrease in $T_c$ (~ 34.8 K) but curiously there is little difference between the $T_c$ of sample D and E although there is an order of magnitude difference in fluence. These results are in contrast with the results of Gandikota et al. that report a systematic decrease in $T_c$ with an increase in normal state resistivity with fluence [16]. We see that the suppression in $T_c$ in $MgB_2$ is much less compared to high $T_c$ ceramic cuprates where empirically it has been established that decrease in $T_c$ ($\delta T_c$) is related to fluence ($\phi$) through the equation $\delta T_c / \phi = 2.9 \times 10^{-11}$ K per ions per $cm^2$ [19]. In the intermetallic $MgB_2$, expectedly, this equation is not followed and the $T_c$ suppression is not linearly related to the fluence. We also note that the transition broadening in all the irradiated samples is less than 1 K. This confirms that the extrinsic defects are uniformly distributed in the sample. The zero-field transition actually becomes sharper after irradiation and this is possibly due to improvement in interfacial matching between the substrate and the film.

Table 1 summarizes the absolute value of resistivity at 40 K, RRR (residual resistivity ratio) and $\Delta\rho$ ($\rho$ (293 K) - $\rho$ (40 K)) for the five samples ascertained from Figure 1. We find that the magnitude of RRR ($\rho$ (293 K) / $\rho$ (40 K)) decreases with irradiation. We also note that $\rho$ (40 K) increases with irradiation dose. This behavior is well supported by other irradiation experiments [15, 16]. Rowell et al. have argued that an increase in $\Delta\rho$ is true measure of post-irradiation decrease in inter grain connectivity [20]. A constant $\Delta\rho$ was interpreted as no change in connectivity up to the fluence of $10^{17}$ ions/$cm^2$ [16]. Our results on the other hand show that while $\Delta\rho$ increases with fluence both for Si and Au irradiation, the effect is much more pronounced for gold ions. Evidently $\Delta\rho$ is larger in dirty limit. To further confirm this trend, plotted in Inset 2 of Figure 1 is the normalized zero field resistivity as a function of temperature. We see that $\rho$ vs. T behavior is more or less similar for the pristine and $Si^{8+}$ irradiated films but it differs for Au irradiated films.

Our results, e.g. small change in $T_c$, significant change in $\Delta\rho$ for the dirty samples and similar $\rho$ vs. T for the unirradiated and Si irradiated samples support the model proposed by Mazin et al. on the two band effects on the transport properties of $MgB_2$ [21]. Following their analysis we conclude that ion irradiation effectively alters the $\pi$ intraband defect scattering. But the question still remains what happens to the $\sigma$ band scattering post irradiation. Between sample D and E we observe that for an order of magnitude difference in dose, the $T_c$ for $Si^{8+}$ irradiation does not change. Between sample B and C on the other hand there is drastic change in $T_c$ when the dose is increased 5 times. As seen in Inset 2 of Figure 1, the $\rho$ vs. T curves for Au irradiation do not superimpose on each other unlike the case with Si irradiation. Clearly the mechanism for $T_c$ suppression with extended defects is different than that for the point defects in $MgB_2$.

We next discuss the transport properties of irradiated samples in the presence of external magnetic field. The upper critical field $H_{c2}$ is a thermodynamic parameter and any modification can only be explained by change in transport mean free path [22]. Plotted in the inset of Figure 2 is the in-field transition for the sample D. Similar measurements were carried out for sample A, B, C, and E. Figure 2 shows $H_{c2}$ versus reduced temperature ($T/T_c$) for all the five samples. It is evident that an increase in $H_{c2}$ is obtained for the Au irradiated sample B and C. For Si irradiated samples, $H_{c2}$ decrease compared to unirradiated samples at high fields. We further note that for the same dose of irradiation while $Au^{15+}$ ions lead to positive curvature in the H-T phase diagram, a linear behavior is observed for $Si^{8+}$ irradiation. Gurevich et al. have suggested that a relatively dirty $\sigma$ band leads to curvature in H- T phase diagram where as a dirtier $\pi$ band leads to linear H- T phase diagram [4]. In this light our data indicate that while irradiation invariably alters the $\pi$ band, extended defects actually alter the $\sigma$ band too.

More support for this is arrived from dρ/dT values at room temperature. For π band scattering Rowell et al. have predicted that the room temperature dρ/dT should be approximately 0.065 μΩ cm/K [23]. From Table 1 for the pristine and Si irradiated samples similar values are obtained. For high dose Au on the other hand dρ/dT is doubled. This again points to the fact that for extended defects σ band properties are vastly altered along with accompanied changes in π band.

To confirm whether the increase in Δρ actually translates into decrease in connectivity and therefore decrease in critical current density, we carried out isothermal magnetization measurements for sample A, C and D. We chose these samples because sample C showed maximum increase in Δρ, where as sample D exhibited minimum increase. Figure 3 shows the magnetization $J_c$ calculated from the magnetization loops using Bean's model [24]. It is interesting to see that critical current density for sample C falls below that of the unirradiated sample A. Sample D on the other hand exhibits an increase in $J_c$ because of combination of two facts a) increase in intragrain pinning and b) negligible decrease in intergrain connectivity. Transport $J_c$ obtained from I –V scans at external field $\mu_0 H = 1$ T also confirmed that inter grain connectivity is severely suppressed in sample C.

As pointed out earlier, the modification of $H_{c2}$ vs. T behavior in a multiband superconductor critically depends on the size and directional properties of the created defects. In Figure 4 we compare the upper critical field as a function of reduced temperature when the field is applied parallel and perpendicular to the ab-plane of the sample. For clarity only the data for sample A, B, and E are shown. In the inset the temperature scans at a constant field of 7 T are plotted. When the field is applied parallel to ab-plane, not much difference in $H_{c2}$ for the three samples is observed. This means that the anisotropy which is defined as $\gamma = H_{c2}^{\parallel} / H_{c2}^{\perp}$ could be higher or lower compared to the pristine sample depending on $H_{c2}^{\perp}$ values. Here $H_{c2}^{\parallel}$ and $H_{c2}^{\perp}$ stand for upper critical field parallel and perpendicular to ab-plane of the sample respectively. From Figure 4 it is very interesting to infer that while the anisotropy decreased with irradiation for extended defects [25], with point defects there is actually a possibility to increase anisotropy. Thus by choosing ions of different energy and atomic number one can in principle tune the scattering mechanism selectively in $MgB_2$.

## 4. Conclusion

To summarize, we have carried out a systematic study of ion-induced defects on superconducting thin films of $MgB_2$. We find that while all defects generally alter the $\pi$ band scattering mechanism, extended defects along the c-axis preferentially affect the $\sigma$ band also. This has significant ramification for $H_{c2} - T$ phase diagram and could lead to an increase in anisotropy after irradiation. The intergrain connectivity is suppressed drastically with extended defects leading to decrease in critical current density. The $dH_{c2} / dT$ slope on the other hand is least affected with isotropic point defects.

**Acknowledgement**

SDK wishes to thank CSIR (India) for financial support. SP acknowledges support form DST (India) under the young scientist scheme and funding for infrastructure development program (FIST).

**Figure captions**

Figure 1 : Temperature dependence of resistivity of $MgB_2$ thin films. Data are shown for unirradiated A (■), Au $1\times10^{12}$ irradiated B (O), Au $5\times10^{12}$ irradiated C (□), Si $5\times10^{12}$ irradiated D (☆), and Si $5\times10^{13}$ irradiated E (Δ). Inset 1 shows the variation in transition temperature. Inset 2 plots normalized resistivity with respect to temperature.

Figure 2 : Upper critical field $H_{c2}$ is plotted against reduced temperature $T/T_c$. The magnetic field is applied perpendicular to ab plane. The joining lines are guides to eye. Inset shows in-field transitions from 0 to 7 T for sample D. The positive curvature for Au irradiated sample is evident. The inset also shows how we have defined $H_{c2}$ for the measurements.

Figure 3 : Variation of magnetization critical current density $J_c$ as a function of field for unirradiated A (■), Au $5\times10^{12}$ irradiated C (□), and Si $5\times10^{12}$ irradiated D (☆) at 5 K and 20 K. The inset shows magnetization loops at T = 5 K. Improvement in $J_c$ is observed only for light ion irradiation.

Figure 4 : Upper critical field $H_{c2}$ is plotted against normalized $T_c$ in perpendicular (closed symbol) and parallel (open symbol) to ab-plane direction for sample A, B, and E. Inset shows superconducting transition with an external field $\mu_0 H = 7$ T. An increase in anisotropy for Si irradiated sample is observed.

# Table 1

| Ion (Sample) Energy (MeV) | Fluence (ions/cm$^2$) | $T_c$ (K) H = 0T | RRR $\rho_{293}/\rho_{40}$ | $\rho_{(40K)}$ ($\mu\Omega$-cm) | $\Delta\rho_{(293-40)}$ ($\mu\Omega$-cm) | $(d\rho/dT)_{293\,K}$ |
|---|---|---|---|---|---|---|
| Unirradiated (A) 0 | 0 | 35.7 | 1.50 | 19.5 | 9.8 | 0.07 |
| Au$^{15+}$ (B) 200 | $1 \times 10^{12}$ | 34.5 | 1.41 | 32.4 | 13.8 | 0.08 |
| Au$^{15+}$ (C) 200 | $5 \times 10^{12}$ | 32.5 | 1.31 | 79.4 | 25.1 | 0.15 |
| Si$^{8+}$ (D) 100 | $5 \times 10^{12}$ | 34.9 | 1.41 | 22.6 | 10.1 | 0.07 |
| Si$^{8+}$ (E) 100 | $5 \times 10^{13}$ | 34.8 | 1.42 | 40.3 | 16.8 | 0.07 |

**Figure 1**

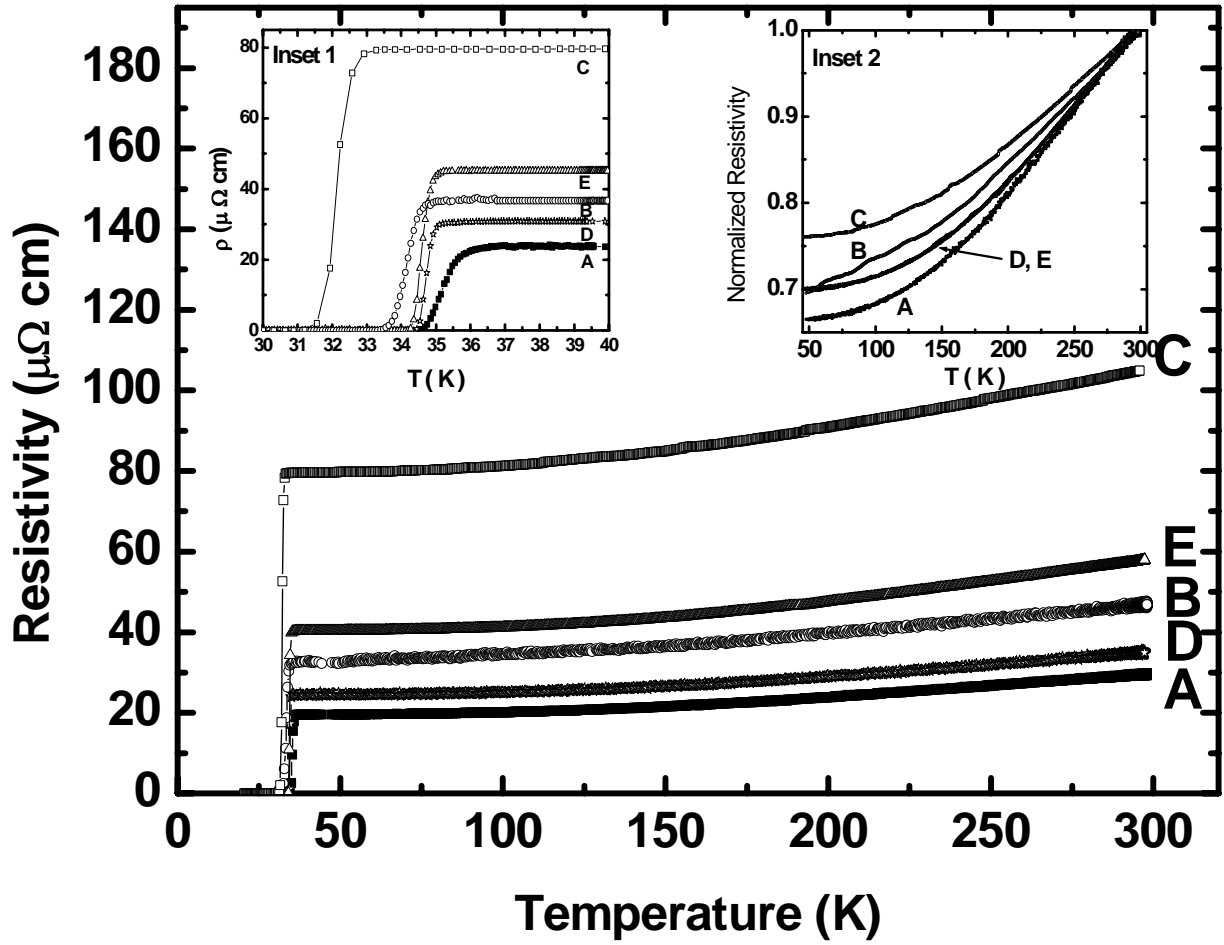

**Figure 2**

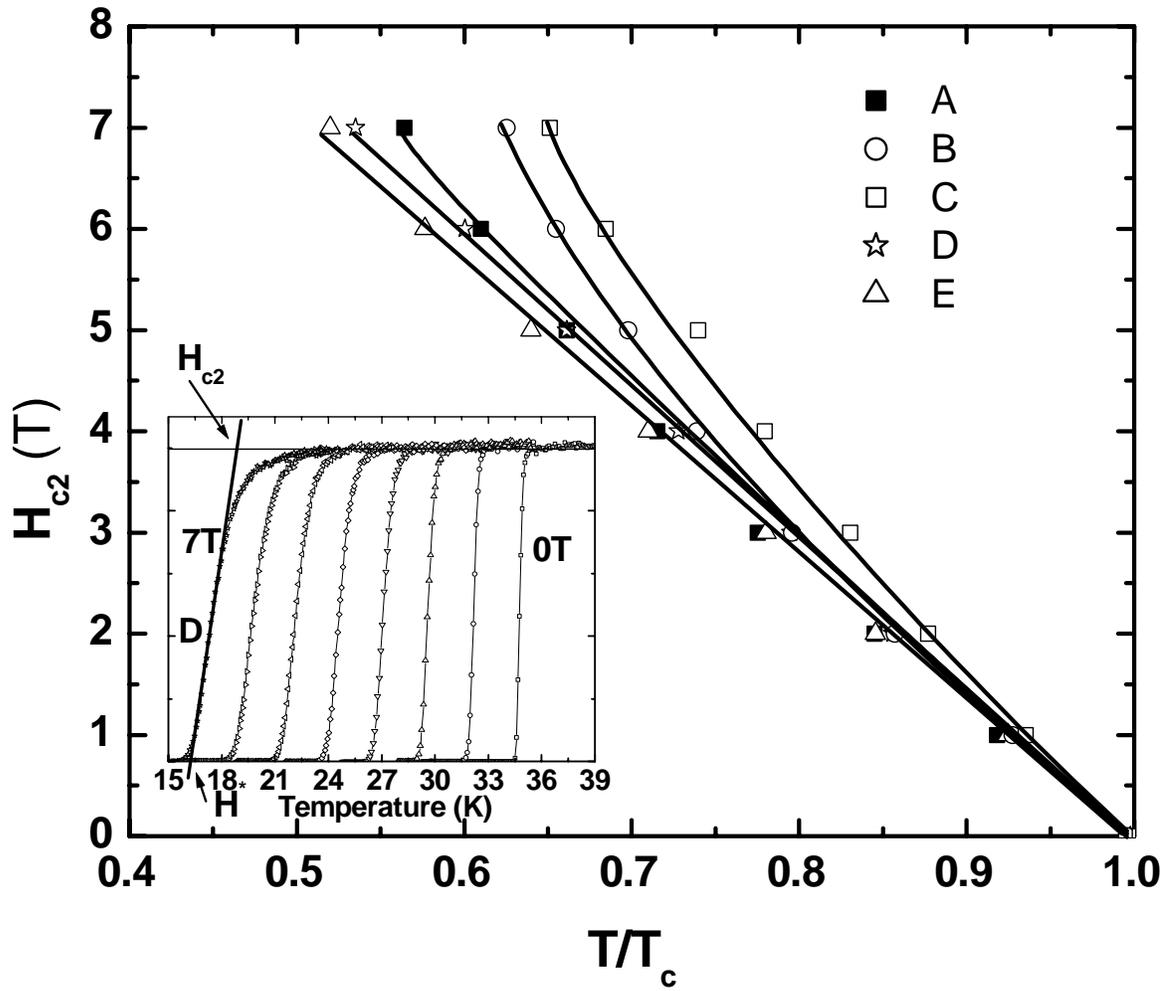

**Figure 3**

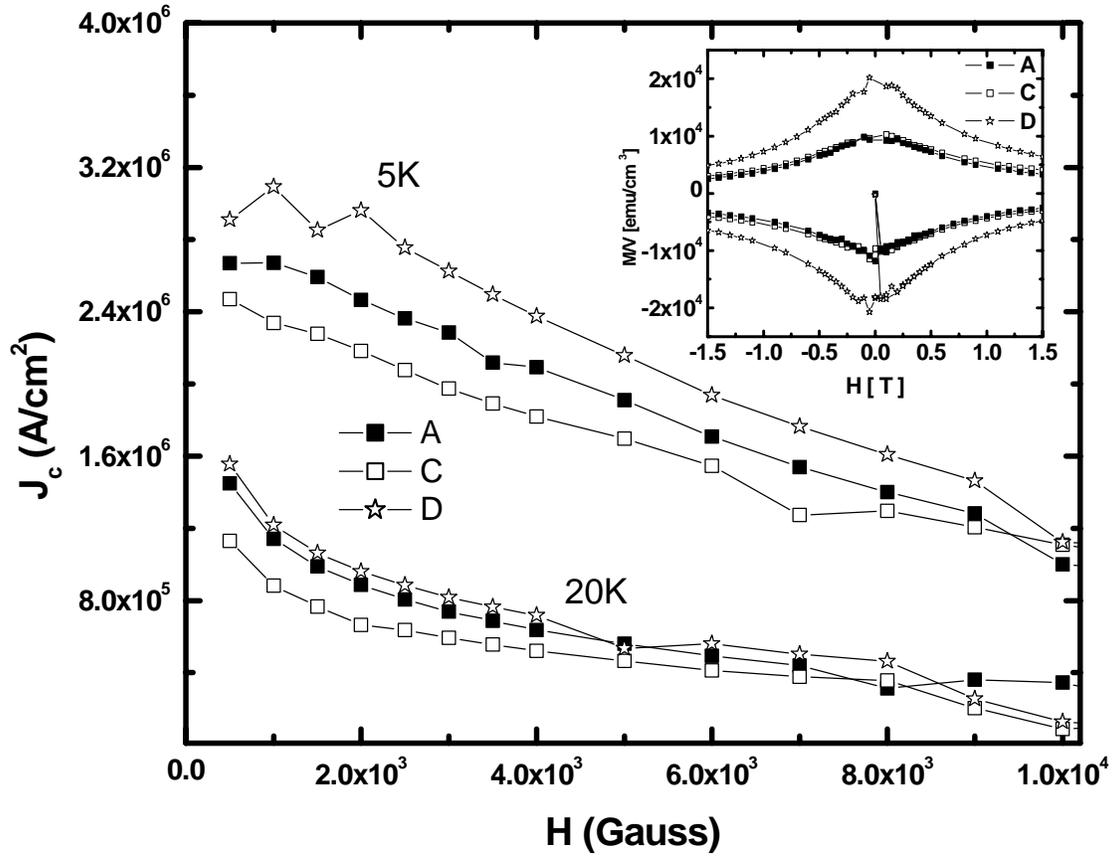

**Figure 4**

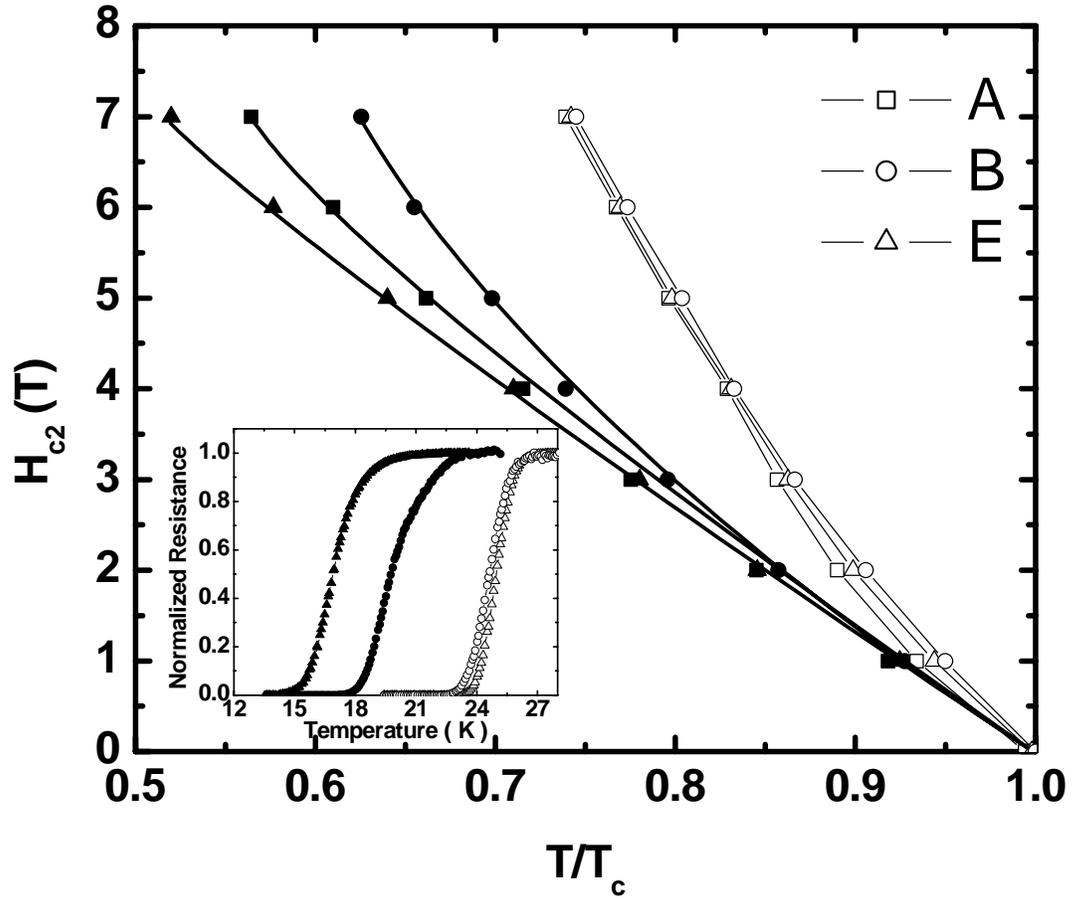